\documentclass[aps,prl,twocolumn,superscriptaddress]{revtex4-2}
\usepackage{graphicx}
\usepackage{acro}
\usepackage{xcolor}

\DeclareAcronym{TMD}{
short = TMD ,
long = Transition-metal dichalcogenide
}
\DeclareAcronym{mdcs}{
short = MDCS ,
long = multidimensional coherent spectroscopy
}
\DeclareAcronym{fwm}{
short = FWM ,
long = four-wave mixing
}
\DeclareAcronym{fwhm}{
short = FWHM ,
long = full-width at half maximum
}
\DeclareAcronym{qise}{
short = QISE ,
long = quantum information science and engineering
}

\begin{document}

%\preprint{}

\title{Inhomogeneous saturation of excitons in monolayer transition-metal dichalcogenides}


\author{Blake T. Hipsley}
\affiliation{Department of Physics, University of Michigan, Ann Arbor, Michigan 48109, USA}
\author{Adam Alfrey}
\altaffiliation{Current Affiliation: Air Force Research Laboratory (AFRL), Wright-Patterson AFB, OH 45433}
\affiliation{Department of Physics, University of Michigan, Ann Arbor, Michigan 48109, USA}

\author{Steven T. Cundiff}
\email{cundiff@umich.edu}
\affiliation{Department of Physics, University of Michigan, Ann Arbor, Michigan 48109, USA}
\affiliation{Quantum Research Institute, University of Michigan, Ann Arbor, Michigan 48109, USA}
%\thanks{}

\date{\today}

\begin{abstract}
We observe that the apparent inhomogeneous broadening, as measured by two-dimensional coherent spectroscopy (2DCS), of the exciton resonance in transition-metal dichalcogenide monolayers depends on the excitation strength. A key strength of 2DCS is the ability to separate inhomogeneous broadening, which primarily contributes to the diagonal linewidth, from homogeneous broadening, which dominates the cross-diagonal linewidth. We show that the fluence dependence of the diagonal linewidth arises from the effective saturation fluence varying with the exciton's resonance energy, i.e., inhomogeneous saturation. These results are critical for interpreting the exciton linewidths, which are often used as a measure of sample quality. 
\end{abstract}

%\keywords{}

\maketitle

\ac{TMD} monolayers have strong light-matter interactions due to the large oscillator strength of the exciton resonance \cite{2014Li_PRB,2018Wang_RMP}, leading to remarkable results such as the demonstration of atomically thin mirrors \cite{2018Back_PRL} and perfect absorbers \cite{2020Horng_PRApplied}. As a consequence, \ac{TMD}s are attracting great interest for optoelectronics and quantum technology applications~\cite{Mak2016-nz,Ren2026-yt}. Successfully exploiting \ac{TMD}s for these applications requires understanding and controlling the processes and phenomena that determine the properties of the excitonic resonances.

One of the most important parameters of a resonance is its linewidth and the nature of the mechanisms determining it, specifically the relative contributions of inhomogeneous broadening versus homogeneous dephasing. In early studies of \ac{TMD} excitons, the exciton resonances were strongly inhomogeneously broadened due to multiple grains, substrate roughness, impurities, and other forms of disorder \cite{Moody2015-linewidthbroadening}. It was discovered that encapsulation in hexagonal boron nitride (hBN), greatly reduced the exciton linewidth \cite{2017Ajayi_2DMat,2017Cadiz_PRX}, which was attributed to the smoother surface provided by hBN, in comparison to silicon dioxide, and passivation of surface states. While there is a clear reduction in the inhomogeneous broadening, partitioning the broadening into inhomogeneous and homogeneous contributions is problematic from a one-dimensional spectrum \cite{2026Alfrey_arXiv}.

\begin{figure}
\includegraphics[width=\linewidth]{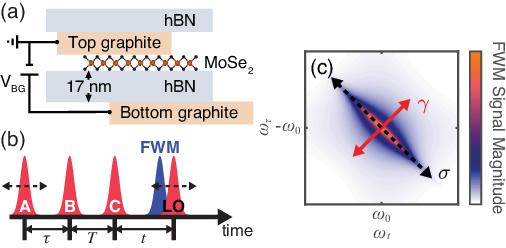}
\caption{(a) Diagram of the hBN-encapsulated MoSe$_2$ monolayer and electrical connections. (b) Schematic of the three-pulse sequence used for MDCS with a local oscillator (LO) for heterodyne detection. (c) Example of a strongly-inhomogeneous 2D spectrum produced by this pulse sequence. For a system with strong inhomogeneity, the on-diagonal linewidth (black dashed) is associated with the inhomogeneous broadening ($\sigma$) and the cross-diagonal linewidth (red solid) is associated with the homogeneous broadening ($\gamma$).\label{fig:sample}}
\end{figure}

Rigorous separation of inhomogeneous broadening and homogeneous dephasing requires the use of more sophisticated spectroscopic techniques, such as \ac{mdcs} \cite{2026Alfrey_arXiv,2023Li_Book}. For the standard \ac{mdcs} two-dimensional (2D) spectrum in the inhomogeneous limit shown in Fig.~\ref{fig:sample}(c), the cross-diagonal width is determined by the homogeneous dephasing rate, $\gamma$, while the diagonal width is due to a combination of $\gamma$ and the width of the inhomogeneous distribution, $\sigma$. Initial applications of \ac{mdcs} to monolayer \ac{TMD}s revealed strong inhomogeneous broadening for unencapsulated flakes \cite{Moody2015-linewidthbroadening,Chen2023}. Even for hBN-encapsulated monolayers, \ac{mdcs} revealed that inhomogeneous broadening was still comparable to the homogeneous dephasing \cite{Martin2020-hbn_encap_linewidth}. The isolation of the homogeneous dephasing also revealed that it depends on excitation strength \cite{Moody2015-linewidthbroadening,Jakubczyk2019-vc-inhomo,Martin2020-hbn_encap_linewidth}, a phenomenon known as excitation-induced dephasing (EID) that had been observed in gallium arsenide quantum wells \cite{Schultheis1986-bl} and that results in new nonlinear signals \cite{1993Wang_PRL,1994Hu_PRB,2006Li_PRL}.

Here we show that, surprisingly, not only does $\gamma$ depend on excitation strength, but that the diagonal width of the 2D spectrum does as well in \ac{TMD} monolayers, specifically MoSe$_2$, that are encapsulated and undoped. This observation is remarkable because in semiconductors, the inhomogeneous broadening is generally attributed to a static disorder landscape, which in \ac{TMD}s arises from impurities, vacancies, adsorbates, substrate roughness, and heterogeneous strain \cite{2019Rhodes_NatMat}, all of which are independent of excitation level. We attribute the change in diagonal linewidth to a variation in the dipole moment across the inhomogeneous distribution, leading to differential saturation. It is essential to account for this variation in the dipole moment when interpreting spectroscopic results and for potential applications of these materials.

Fig.~\ref{fig:sample}(a) shows a diagram of the sample. Exfoliated MoSe$_2$ flakes always have some degree of intrinsic doping due to impurities and the substrate \cite{2017Jadczak_Nanotech}.  We include top and bottom graphite contacts that allow for electrostatic depletion of residual carriers due to intrinsic doping, as demonstrated by the reduction in the trion/Fermi polaron peak from -2.4 to 0.4 V in the linear reflection contrast spectrum \cite{supp}. All data presented here are taken at a back gate voltage of 0.0 V where the intrinsic carriers are depleted. The \ac{mdcs} pulse scheme used for the spectra here is shown in Fig.~\ref{fig:sample}(b). In \ac{mdcs}, three pulses (denoted as A, B, and C) interact within a material to generate a \ac{fwm} signal, $E_S \propto E_A^* E_B E_C$, which is interfered with a fourth pulse (denoted as LO) acting as a local oscillator for coherent heterodyne detection.  The four pulses are separated by time delays $\tau$, $T$, and $t$, respectively. Pulses A and LO are scanned to vary the time delays $\tau$ and $t$, respectively. Fourier transforming with respect to $\tau$ and $t$ produces a 2D spectrum with axes $\omega_\tau$ and $\omega_t$ (Fig.~\ref{fig:sample}(c)). Since the electric field of pulse A is conjugated, $\omega_\tau < 0$, and the diagonal corresponding to $|\omega_\tau|=\omega_t$ goes from the top left to the lower right of the plot. For single-quantum spectroscopy, the electronic structure between states one quanta of energy apart is probed. The laser spectrum has a \ac{fwhm} of approximately 89 meV~\cite{supp}, which is much wider than any spectral line and thus distortion of the spectrum due to a finite laser bandwidth can be neglected. All beams are co-linearly polarized.

\begin{figure}\centering
\includegraphics[width=\linewidth]{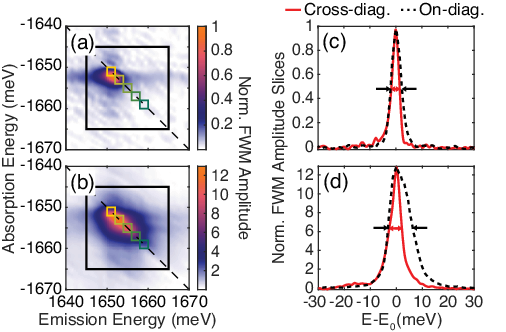}
\caption{Representative single-quantum rephasing spectra for an undoped (0 V bias) MoSe$_2$ monolayer at 7.1 K and a population delay of $T = 100$~fs. Spectra were measured with excitation fluences of (a) 2.49 and (b) 74.6 $\mu$J/cm$^2$ per beam. Included are outlines for the areas integrated over to determine the integrated FWM signals in Figs.~\ref{fig:XINT_vs_XDensity}(a) and~\ref{fig:XINT_vs_XDensity}(b). (c) and (d) Cross- (red solid) and on-diagonal (black dashed) slices of the corresponding single-quantum rephasing spectra with arrows indicating the full-width at half-maximum for each. $E_0$ is the peak energy.\label{fig:MDCS_spectra}}
\end{figure}

The 2D spectrum allows the homogeneous ($\gamma$) and inhomogeneous linewidths ($\sigma$) to be extracted, where in highly inhomogeneously broadened samples the on-diagonal is associated with inhomogeneous linewidth and the cross-diagonal is associated with the homogeneous linewidth~\cite{Siemens-linewidths}. In samples of lower inhomogeneity (i.e., $\sigma~\approx~\gamma$), homogeneous and inhomogeneous broadening effects contribute to both slices and must be separated by simultaneous fitting both~\cite{Siemens-linewidths}, assuming a third-order optical response. As we will describe later, this assumption of a third-order response does not apply to all of the data we report here; thus, we use the \ac{fwhm} values rather than fitting to extract $\sigma$ and $\gamma$. To obtain robust estimates of the FWHM, we use local fits as described in the supplement materials \cite{supp}.

Figs.~\ref{fig:MDCS_spectra}(a) and \ref{fig:MDCS_spectra}(b) show single-quantum rephasing spectra taken on the undoped MoSe$_2$ monolayer at low and high excitation fluences of 2.49 and 74.6 $\mu$J/cm$^2$ per beam, respectively (all fluences quoted hereafter are per beam). Figs.~\ref{fig:MDCS_spectra}(c) and \ref{fig:MDCS_spectra}(d) show the corresponding on- and cross-diagonal slices of the spectra with arrows indicating the \ac{fwhm}s of each. The low fluence spectrum peaks at 1652 meV, corresponding to the resonance of the MoSe$_2$ exciton. Increasing the fluence causes a small shift of the peak to higher energy by 0.6 meV. More notably, at the higher excitation fluence, the peak becomes elongated along the diagonal, more than doubling the on-diagonal \ac{fwhm} of 4.41 meV to 9.48 meV. For comparison, the cross-diagonal \ac{fwhm} only changed from 3.25 meV to 4.84 meV, suggesting the resonance has become more inhomogeneously broadened with increasing fluence.

\begin{figure}
\includegraphics[width=\linewidth]{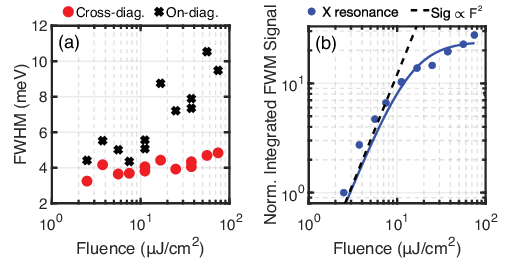}
\caption{(a) Full-width at half maxima for the cross- and on-diagonal slices of the single-quantum rephasing spectra as a function of incident fluence in MoSe$_2$ at 0 V bias. (b) Signal strength for the exciton resonance integrated from 1645 to 1665 meV (large black squares in Figs.~\ref{fig:MDCS_spectra}(a) and~\ref{fig:MDCS_spectra}(b)) along both the absorption and emission axes, as a function of fluence per beam on MoSe$_2$ at 0 V bias normalized to the integrated signal at a fluence of 2.49~$\mu$J/cm$^2$. The black dashed line represents how the integrated signal should grow if it follows a third-order optical response, along with a fit of the integrated signal to Eq.~(\ref{eq:saturation_tot}) (solid blue line).\label{fig:LWs_vs_XDensity}}
\end{figure}

The increase in the linewidths is verified by plotting the \ac{fwhm} of diagonal and cross-diagonal slices of the single-quantum rephasing spectra as a function of excitation fluence in Fig.~\ref{fig:LWs_vs_XDensity}(a). The elongation of the spectrum along the diagonal corresponds with the increase in the on-diagonal \ac{fwhm}. As the exciton resonance shifts with increasing fluence, the positions of the cross-diagonal slices shift slightly to higher energies due to the changing diagonal lineshape.

\begin{figure*}
\includegraphics[width=\linewidth]{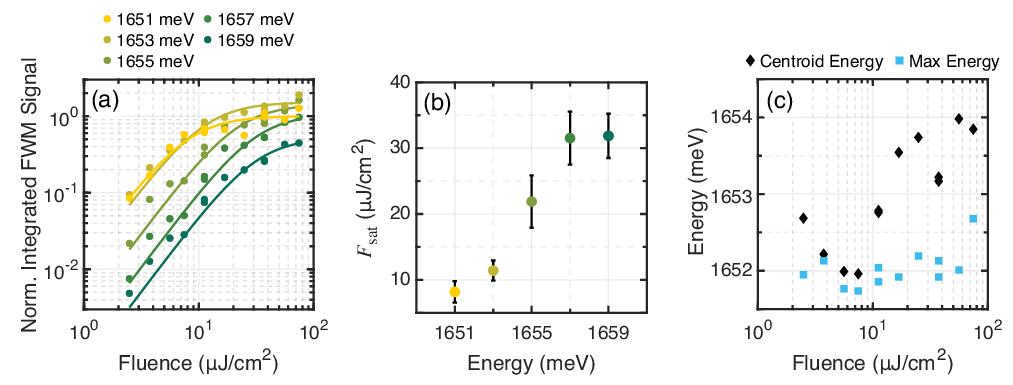}
\caption{(a) Integrated FWM signal for different regions of the exciton resonance as a function of fluence. The integrated signal strength was determined by integrating the single-quantum rephasing spectrum from $\pm~1$ meV about the labeled energy along both the absorption and emission axes, and normalizing to the FWM signal measured at a fluence of 2.49~$\mu$J/cm$^2$, integrated over 1645 to 1665 meV along both axes. These integration areas are indicated by colored boxes on Figs.~\ref{fig:MDCS_spectra}(a) and~\ref{fig:MDCS_spectra}(b). Provided for each energy is a fit of the saturation behavior of the integrated signals. (b) Effective saturation fluences determined by fits of the integrated FWM signal to Eq.~(\ref{eq:saturation_tot}). Errors determined using the covariance matrix of the fits. (c) Energies of the centroid and maximum of single-quantum rephasing spectra plotted as a function of excitation fluence.
\label{fig:XINT_vs_XDensity}}
\end{figure*}

In the standard framework for interpreting a \ac{mdcs} resonance lineshape using a third-order perturbation expansion ($\chi^{(3)}$ regime) of the optical Bloch equations \cite{Siemens-linewidths}, these results would be interpreted as showing that the width of the inhomogeneous distribution of exciton states increases with increasing exciton population, which is governed by the excitation fluence. To determine if this is the only interpretation, the validity of the underlying assumptions must be checked. Specifically, we determine if the approximation of a third-order response is valid by plotting the strength of the spectrally integrated peak strength as a function of excitation fluence in Fig.~\ref{fig:LWs_vs_XDensity}(b). To ensure the entire resonance is integrated over, even at high fluences when the spectra become elongated, each single-quantum rephasing spectrum is integrated along the absorption and emission energies from 1645 meV to 1665 meV, as indicated by the large black squares in Figs.~\ref{fig:MDCS_spectra}(a) and~\ref{fig:MDCS_spectra}(b). In this measurement, we vary the strength of all four beams, the excitation beams plus the local oscillator beam, $E_{LO}$. Since the signal is measured using heterodyne detection, the strength of the measured signal is $S \propto E_S E_{LO}^* \propto E_A^* E_B E_C E_{LO}^* \propto I^2$, where $I$ is the intensity of the laser at the sample. Since the fluence is proportional to the intensity, this dependence also applies to fluence. The dashed line in Fig.~\ref{fig:LWs_vs_XDensity}(b) shows this expected dependence. The measured data deviate from the expected dependence, showing that the assumption of a third-order response fails as the signal saturates at high fluences. To model this saturation behavior, we start with the expression for determining the saturation of the gain coefficient of a two-level system as a function of intensity and adapt it to account for the fact that the spectrally integrated signal, $S$, should grow quadratically with fluence in the low-fluence limit and approach a constant value in the high-fluence limit to be
\begin{equation}
    S(F) = A\frac{F^2}{1+\left(F/F_\textrm{sat}\right)^2},
    \label{eq:saturation_tot}
\end{equation}
where $F$ is the fluence, $A$ is the amplitude, and $F_\textrm{sat}$ is the effective saturation fluence (see the supplemental material \cite{supp}). Fitting the integrated resonance data in Fig.~\ref{fig:LWs_vs_XDensity}(b) to Eq.~(\ref{eq:saturation_tot}), we obtain $F_\textrm{sat} = 14~\pm~2 ~\mu$J/cm$^2$.

The observation that the measurements are not entirely within the $\chi^{(3)}$ regime indicates that we must consider other mechanisms for the increase in the diagonal linewidth. One possibility is that the oscillator strength varies across the inhomogeneous distribution, leading to photon-energy--dependent saturation intensity/fluence. Indeed, a variation in the oscillator strength across the inhomogeneous distribution was predicted for excitons in a disordered quantum well \cite{Rohan_Localized_excitons}. To check for this possibility, the integrated signals in $\pm$ 1 meV regions of the single-quantum rephasing spectra within the exciton resonance are shown in Fig.~\ref{fig:XINT_vs_XDensity}(a). We generalize Eq.~(\ref{eq:saturation_tot}) to allow for an energy dependence of the amplitude, $A \rightarrow A(E)$, and effective saturation fluence $F_{sat} \rightarrow F_{sat}(E)$ and fit each saturation curve. Clearly, as quantified in Fig.~\ref{fig:XINT_vs_XDensity}(b), $F_{sat}(E)$ increases with energy. Consequently, the strength of the high-energy side of the inhomogeneous distribution increases relative to the low-energy side with increasing fluence, thereby causing the diagonal elongation. Naturally, we would also expect a shift in the apparent resonance energy with increasing fluence. The data in Figs.~\ref{fig:MDCS_spectra}(a) and~\ref{fig:MDCS_spectra}(b) suggested that the peak does not shift significantly; however, the elongation to higher energy with increasing fluence is clearly visible in Figs.~\ref{fig:MDCS_spectra}(c) and~\ref{fig:MDCS_spectra}(d). To quantify these observations, in Fig.~\ref{fig:XINT_vs_XDensity}(c), we plot both the peak position and centroid as a function of fluence. The shifting of the centroid to higher energy is clearly visible. The centroid is more sensitive because the lower-energy states do not decrease; rather, they continue to increase, albeit more slowly than the higher-energy states.

Excitons in \ac{TMD}s exhibit many density-dependent effects that must be considered as alternative explanations for our observations. Density-dependent blue shifts due to exciton-exciton interactions or phase-space filling would cause the peak to shift off the diagonal, as the blue shift would be different during the time periods $t$ and $\tau$, which we do not observe. EID would broaden the cross-diagonal linewidth more strongly than the cross-diagonal linewidth, which is inconsistent with our results. While biexcitonic effects cannot be ruled out, they give rise to peaks below the diagonal \cite{2009Bristow_PRB,2017Hao_NComms} and thus do not explain our observations. In addition, we can ignore the effects of surpassing the exciton Mott density as we estimate a maximum excitation density of $2.82\times10^{12}$~excitons/cm$^2$, far below the estimated Mott density of $3.18\times10^{13}$~excitons/cm$^2$(see supplemental material~\cite{supp}).

TMDs are notorious for sample-to-sample variations in their properties. To confirm that our observations are not unique to this sample, in the supplemental material we include data from a second hBN-encapsulated MoSe$_2$ flake that is not gated~\cite{supp}. This sample displays much stronger EID, which makes the effect less clear, but nevertheless, there is a clear increase in the diagonal linewidth and shift of the centroid to higher energies.

The results presented here clearly show that the diagonal linewidth of the 2D coherent spectra of excitons in MoSe$_2$ can depend on the excitation fluence. These results can be explained if inhomogeneous broadening arises from disorder that localizes excitons. Previous simulations have shown that, for excitons localized by energetic fluctuations, the oscillator strength decreases from low energy to high energy within the inhomogeneous distribution, while the density of states increases \cite{Rohan_Localized_excitons}. The predicted dependencies would produce the fluence-dependent diagonal elongation that we observe. Moreover, the energy-dependent $F_{sat}(E)$ shown in Fig.~\ref{fig:XINT_vs_XDensity}(b) are consistent with low-energy, high-oscillator strength localized excitons and high-energy, lower-oscillator delocalized excitons. Thus, these experimental results support the idea that disorder localizes excitons and plays an essential role in determining their nonlinear optical response. A microscopic theory for these results is beyond the scope of this experimental work, as it will require including the inhomogeneous broadening in more detail than simply convolving the homogeneous results with a static inhomogeneous distribution \cite{2016Selig_NComms}.

Beyond the physical insight into the nature of the excitonic nonlinear optical response, these results are also critical for interpreting a broad range of spectroscopic results. Most directly, they show that reports of inhomogeneous linewidths should specify whether, and how, the linewidths depend on excitation fluence. The relevance of the high-fluence versus low-fluence results will depend on the nature of the experiment or application. Furthermore, the results show that the density-dependent shifts of the exciton resonance may not be solely due to interactions, which is a common interpretation, but may also be due to the inhomogeneous saturation that we have revealed here. The phenomenon of inhomogeneous saturation is likely to occur in materials other than TMDs, our results have a broad relevance to many systems with a strong nonlinear optical response.

\begin{acknowledgments}
\textit{Acknowledgments} -- We acknowledge funding from DOE/BES Grant DE-SC0022179 and NSF Grant 2004286.
\end{acknowledgments}

\textit{Data availability} -- There are no publicly available research data or software supporting this manuscript. Requests for further information or data should be sent to the authors.

\textit{Conflict of interest} --  STC is a co-founder of MONSTR Sense Technologies LLC, which manufactures co-linear MDCS Spectrometers.

\bibliography{references.bib}

\end{document}